\documentclass[aps,prl,preprint,amsfonts,nobibnotes,superscriptaddress,showpacs]{revtex4}

\usepackage{dcolumn}% Align table columns on decimal point
\usepackage{amsmath}
\usepackage{amssymb}
\usepackage{graphicx}
\usepackage{bm}
\usepackage{upgreek,xspace}
\usepackage{color}
\usepackage{titlesec}

\begin{document}

\title{Phonon-Driven Selective Modulation of Exciton Oscillator Strengths in Anatase TiO$_2$ Nanoparticles}

\vspace{2cm}

\author{Edoardo Baldini}
	\affiliation{Department of Physics, Massachusetts Institute of Technology, Cambridge, Massachusetts, 02139, USA}
	\affiliation{Laboratory of Ultrafast Spectroscopy, ISIC and Lausanne Centre for Ultrafast Science (LACUS), \'Ecole Polytechnique F\'ed\'erale de Lausanne (EPFL), CH-1015 Lausanne, Switzerland}
	
	\author{Tania Palmieri}
	\affiliation{Laboratory of Ultrafast Spectroscopy, ISIC and Lausanne Centre for Ultrafast Science (LACUS), \'Ecole Polytechnique F\'ed\'erale de Lausanne (EPFL), CH-1015 Lausanne, Switzerland}

\author{Adriel Dominguez}
	\affiliation{Departamento Fisica de Materiales, Universidad del Pa\'is Vasco, Av. Tolosa 72, E-20018, San Sebastian, Spain}
	
\author{Pascal Ruello}
\affiliation{Institut des Mol\'ecules et Mat\'eriaux du Mans, UMR CNRS 6283, Le Mans Universit\'e, 72085 Le Mans, France}

\author{Angel Rubio}
	\affiliation{Departamento Fisica de Materiales, Universidad del Pa\'is Vasco, Av. Tolosa 72, E-20018, San Sebastian, Spain}
	\affiliation{Max Planck Institute for the Structure and Dynamics of Matter, Hamburg, Germany}

\author{Majed Chergui}
	\affiliation{Laboratory of Ultrafast Spectroscopy, ISIC and Lausanne Centre for Ultrafast Science (LACUS), \'Ecole Polytechnique F\'ed\'erale de Lausanne (EPFL), CH-1015 Lausanne, Switzerland}

\date{\today}

\begin{abstract}
The way nuclear motion affects electronic responses has become a very hot topic in materials science. Coherent acoustic phonons can dynamically modify optical, magnetic and mechanical properties at ultrasonic frequencies, with promising applications as sensors and transducers. Here, by means of ultrafast broadband deep-ultraviolet spectroscopy, we demonstrate that coherent acoustic phonons confined in anatase TiO$_2$ nanoparticles can selectively modulate the oscillator strength of the two-dimensional bound excitons supported by the material. We use many-body perturbation-theory calculations to reveal that the deformation potential is the mechanism behind the generation of the observed coherent acoustic wavepackets. Our results offer a route to manipulate and dynamically tune the properties of excitons in the deep-ultraviolet at room temperature.
\end{abstract}

\pacs{}
%\keywords{}

\maketitle

\section{Introduction}

The field of excitonics is attracting increasing interest for storing, guiding and manipulating light at the nanoscale via collective bound states of electron-hole pairs. As such, one of the primary goals of research is to reach high control and tunability of the exciton optical properties (i.e. linewidth, energy, oscillator strength) in distinct spectral ranges through different external perturbations. Typical examples of control parameters include temperature, pressure, electric/magnetic field, mechanical stress and photoexcitation \cite{miller1984band, kuo2005strong, stier2016exciton}. On the one hand, varying the exciton peak energy is highly desirable for ensuring wavelength tunability to exciton radiative recombination in direct gap semiconductors, or for discriminating between degenerate exciton states in valleytronic materials \cite{aivazian2015magnetic, srivastava2015valley, macneill2015breaking}. On the other hand, modifying the exciton oscillator strength has led to considerable developments in low-power all-optical switching and signal processing, with the photo-induced saturation of the excitonic absorption being the most significant one \cite{miller1981dynamic}.

A novel frontier in the field involves pushing these effects to a coherent regime through the use of tailored ultrashort laser pulses, realizing a time-periodic modulation of the exciton optical properties \cite{baldini_book}. Recently, huge efforts have been devoted to dynamically modulate the peak energy of excitons \cite{kim2014ultrafast, sie2015valley, akimov2006ultrafast, scherbakov2007chirping, sagar2008size}. A promising framework is based on the optical Stark effect, in which off-resonant (i.e. below gap) photons dress the excitonic states of a material within the strong coupling regime. With this method, remarkable exciton shifts of 10-18 meV have been obtained in monolayer transition metal dichalcogenides at room temperature (RT) \cite{kim2014ultrafast, sie2015valley}. An alternative approach relies on tuning the exciton energy by photoinducing a time-dependent ionic motion. In this scenario, exciton shifts as large as 10 meV have been demonstrated in semiconductor nanostructures, although at very low temperatures \cite{akimov2006ultrafast, scherbakov2007chirping}. Experiments at RT are rarer, due to the scarcity of materials in which excitons can still emerge as robust quasiparticles and coherent phonons are not suppressed by fast thermal dephasing. A prototypical example is offered by quantum dots, where the exciton-phonon coupling is enhanced by their low dimensionality, leading to exciton shifts of several meV \cite{sagar2008size}.

However, so far, achieving the selective modulation of an exciton oscillator strength has remained a challenge. In this scenario, only the exciton peak amplitude is expected to periodically react to the external coherent perturbation, while the resonance energy remains anchored to its equilibrium value. To our knowledge, this effect has only been reported for charge-transfer excitons of specific molecular systems \cite{kano2002observation}, with no direct counterparts in inorganic materials. Reaching a coherent amplitude modulation of an exciton resonance would allow a range of potential applications, as the exciton oscillator strength can act as a sensor of the acoustic field.

In this Letter, we demonstrate the phonon-driven selective modulation of the exciton oscillator strength in the technologically-relevant oxide anatase TiO$_2$. This semiconductor crystallizes in a tetragonal unit cell built on a network of corner- or edge-sharing TiO$_6$ octahedra, with a substantial difference between the lattice constants a = 3.78 \AA~ and c = 9.51 \AA. In its nanosized form, it is widely recognized as a superior system for several applications, ranging from photocatalysis \cite{fujishima1972electrochemical} to dye-sensitized solar cells \cite{ref:oregan} and sensors \cite{bai2014titanium, he2017micro}. In a previous work \cite{ref:baldini_TiO2}, we demonstrated that pronounced electron-hole correlations govern the deep-ultraviolet (UV) optical properties of this material, giving rise to strongly bound excitons with a binding energy exceeding 150 meV \cite{ref:baldini_TiO2}. The lowest-energy (a-axis) exciton is dipole-allowed when the light polarization lies along the (001) crystallographic plane, and possesses a two-dimensional wavefunction in the three-dimensional crystal lattice (Fig. 1(a)). The highest-energy (c-axis) exciton is active for light polarized along the [001] axis, and is characterized by a localized and almost spherical wavefunction (Fig. 1(b)). With respect to these bound states, conventional nanosystems such as TiO$_2$ nanoparticles (NPs) with diameters $>$ 5 nm behave similarly to the bulk, as they are larger than the exciton Bohr radius (3.2 nm). In addition, excitons are notably very robust against temperature and sample quality, clearly appearing in the defect-rich NPs and films used in RT applications \cite{ref:baldini_TiO2}. Importantly, these excitons also possess a strong coupling to the lattice degrees of freedom \cite{ref:baldini_TiO2, ref:tang_urbach, baldini_rutile}, making them ideal candidates to test schemes of dynamical manipulation based on a photoinduced ionic motion. In this respect, nanosized systems offer a high degree of flexibility with regard to coherent acoustic phonons, as the latter become confined entities when the phonon coherence length is comparable to the size of the system \cite{ref:hodak, ref:delfatti, ref:krauss, ref:verma, ref:chern, ref:rossi}. Under these conditions, different geometries of the nano-objects lead to a discrete set of eigenmodes that can be tested for ultrafast exciton control.

\begin{figure}[tb]
	\begin{center}
		\centering
		\includegraphics[width=\columnwidth]{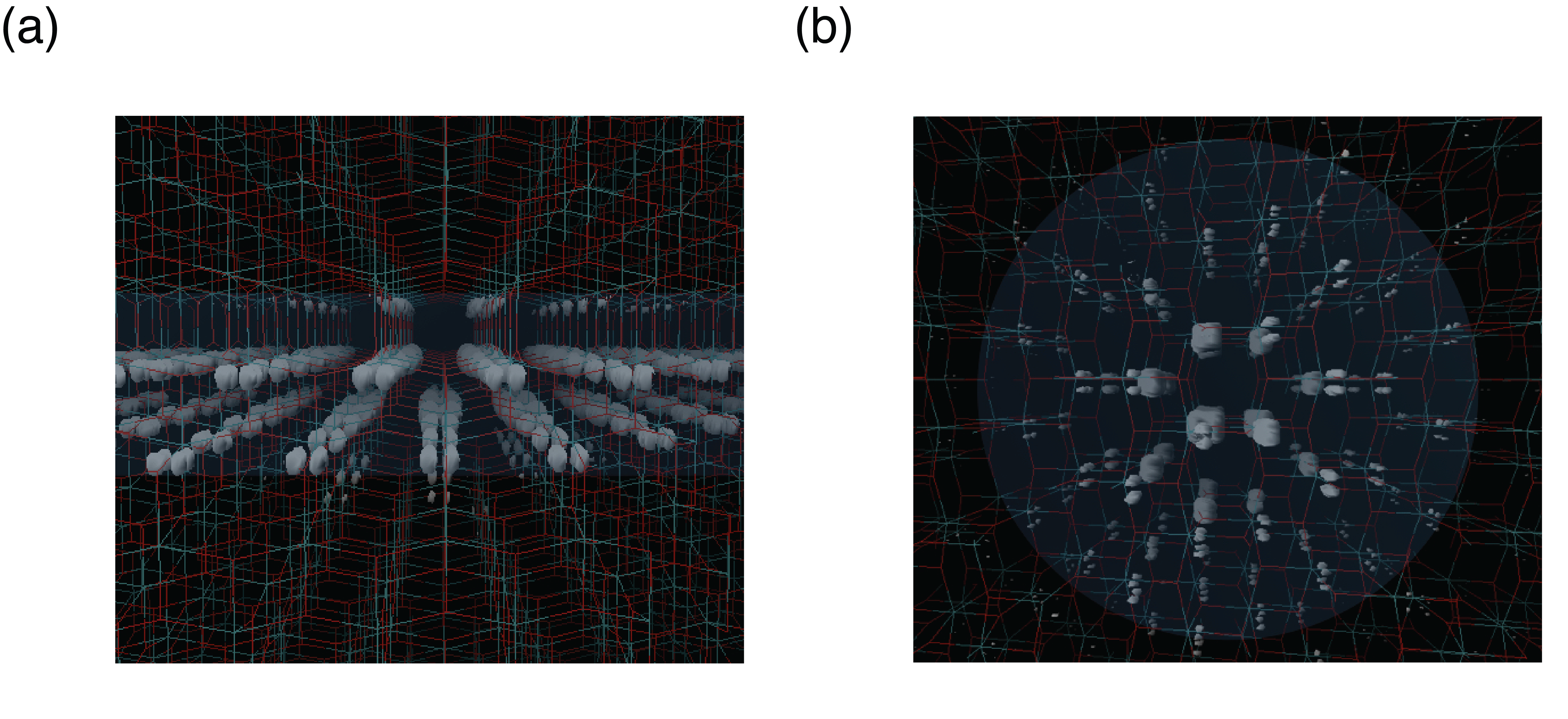}
		\caption{Isosurface representation of the electronic configuration when the hole of the considered excitonic pair is localized close to one oxygen atom. The grey region represents the excitonic squared modulus wavefunction. (a) Bound a-axis exciton. (b) Bound c-axis exciton.}
		\label{fig:1}
	\end{center}
\end{figure}

Here, we use an ultrashort deep-UV laser pulse to impulsively launch confined coherent acoustic phonons inside spherical anatase TiO$_2$ NPs. By means of ultrafast broadband deep-UV spectroscopy, we observe that the resulting mechanical motion of the NPs induces a giant coherent modulation of the two-dimensional exciton oscillator strength, without affecting the peak energy of the resonance. To rationalize the generation mechanism of the coherent strain pulses, we combine our results with advanced many-body perturbation theory calculations. These findings shed light on strong exciton-phonon coupling phenomena in nanosized anatase TiO$_2$, and pave the way to exploiting the material's two-dimensional exciton as a promising basis for new pressure sensors.

Anatase TiO$_2$ NPs with an average diameter of 25 nm were prepared using the sol-gel method \cite{ref:mahshid}, dispersed in a colloidal solution and circulated into a quartz flow-cell at RT. For the ultrafast broadband deep-UV experiments, we used the set-up described in Ref. \cite{ref:aubock}. Specifically, the NPs were excited with a deep-UV pump pulse with photon energy above the a-axis exciton peak, thus generating uncorrelated electron-hole pairs. Subsequently, the relative changes in the NPs absorption ($\Delta$A) were monitored over a broad spectral range covering the exciton feature (3.40 - 4.50 eV). The time resolution was 150 fs. Many-body perturbation theory at the GW level \cite{ref:hedin1, ref:onida} was employed to compute the band structure of the material for an unstrained unit cell and in the presence of hydrostatic pressure. More details are provided in the Methods section.

\begin{figure*}[tb]
	\begin{center}
		\centering
		\includegraphics[width=1\columnwidth]{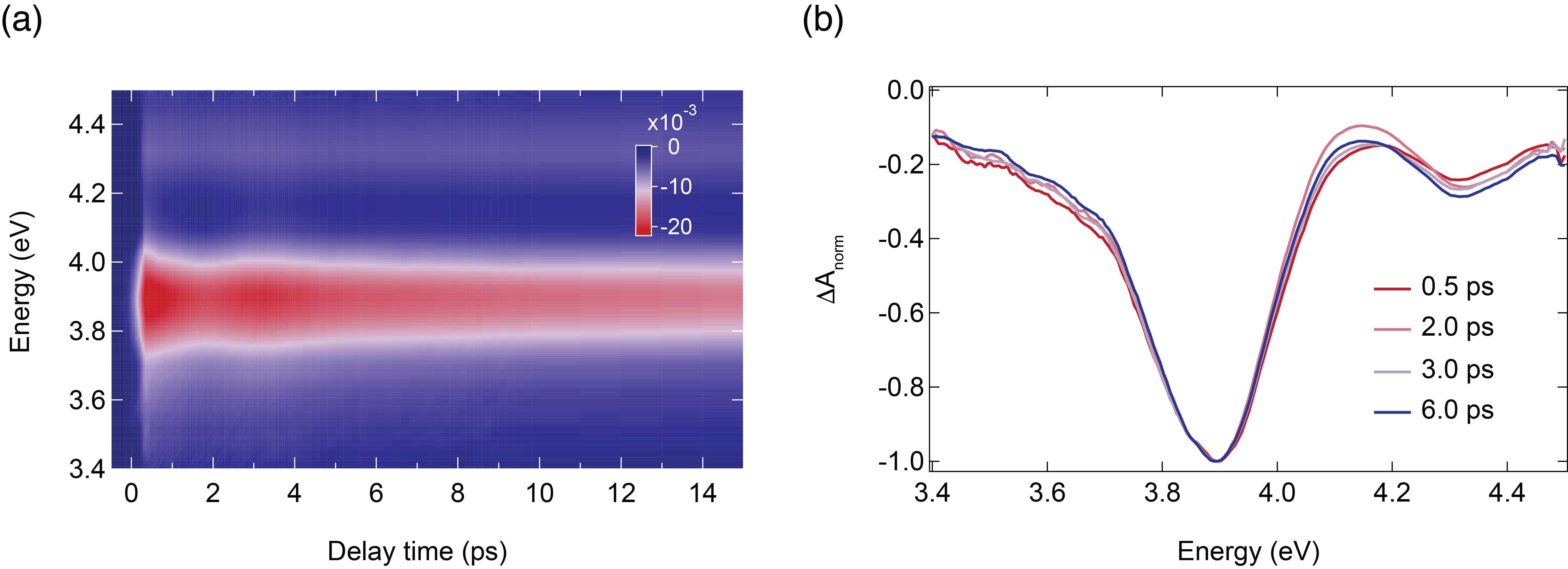}
		\caption{(a) Color map of $\Delta$A measured on a colloidal solution of anatase TiO$_2$ NPs (average diameter of 25 nm) as a function of probe photon energy and time delay between pump and probe. The time resolution is estimated to be 150 fs, the pump photon energy is set at 4.05 eV, and the photoexcited carrier density is $n$ = 5.7 $\times$ 10$^{19}$ cm$^{-3}$. (b) $\Delta$A spectra as a function of probe photon energy at representative delay times between pump and probe.}
		\label{fig:AcPh_2D_MAP_negative}
	\end{center}
\end{figure*}

Figure 2(a) displays the color map of $\Delta$A as a function of probe photon energy and time delay between pump and probe. The photoexcited carrier density is $n\sim$ 5.7 $\times$ 10$^{19}$ cm$^{-3}$. Although the signal was measured up to 1 ns, the map purposely zooms into the first 15 ps. The signal is characterized by a negative response over the entire probe range, and it exhibits two long-lived features around 3.88 eV and 4.35 eV. These structures correspond to the a- and c-axis bound excitons, respectively \cite{ref:baldini_TiO2}, which are bleached as a result of above-gap photoexcitation (as discussed in detail in Ref. \cite{ref:baldini_excitonbleach}). Contrary to the transient reflectivity measurements in single crystals \cite{ref:baldini_TiO2}, both excitons are observed in the ultrafast response of NPs due to their random orientation in solution. Direct inspection of Fig. 2(a) reveals a prominent coherent oscillation at the a-axis exciton band in the first few ps. In the following, we focus on this feature, which indicates the presence of one (or more) collective mode(s) within the photoexcited NPs. To this aim, we first investigate how the exciton spectrum evolves as a function of time during the emergence of the oscillation. Normalized $\Delta$A spectra are shown in Fig. 2(b) at representative time delays. We observe that the a-axis exciton lineshape undergoes a change of asymmetry over time, while maintaining the same peak energy. 
%Fitting the $\Delta$A spectra with a Fano profile over the 3.40 - 4.10 eV energy range and 0.3 - 6 ps time window allows us to quantitatively describe the time evolution of the relevant exciton properties, as shown in Fig. 2(c,d). We find that the coherent modulation only renormalizes the exciton oscillator strength, while the exciton peak energy remains anchored at 3.88 eV for all time delays (Fig. 2(c)). The evolution of the exciton linewidth and asymmetry is also affected by the coherent modulation, resulting in an anti-phase oscillation between the two quantities (Fig. 2(d)).

\begin{figure*}[tb]
	\begin{center}
		\centering
		\includegraphics[width=\columnwidth]{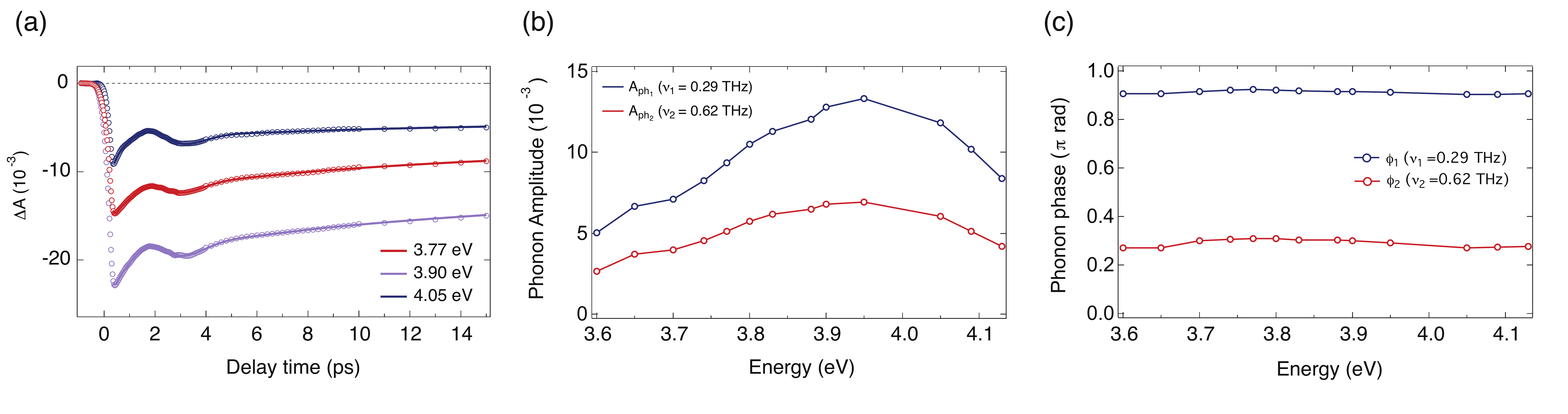}
		\caption{(a) Experimental time traces of $\Delta$A for different probe photon energies (dotted lines) and results of the global fit analysis (solid lines). (b) Contribution to the $\Delta$A response of the two oscillatory components obtained from the global fit analysis. (c) Probe photon energy-dependence of the phase of the oscillations.}
		\label{fig:AcPh_DAS}
	\end{center}
\end{figure*}

Complementary information on the dynamics is obtained from the analysis of the temporal traces at representative photon energies on the red wing (3.77 eV), the centre (3.90 eV) and the blue wing (4.05 eV) of the a-axis exciton peak (Fig. 3(a)). They all exhibit a resolution-limited rise of the bleach signal, followed by a long-lived exponential decay persisting beyond 1 ns \cite{ref:baldini_excitonbleach, baldini2017interfacial} on top of which the giant low-frequency oscillation clearly emerges and undergoes a dramatic damping after one period (with the same frequency) at all probe photon energies. More importantly, all temporal traces across the exciton peak show oscillations \textit{in phase}, an aspect that is at odds with the behavior expected for an exciton coupled to a coherent collective mode \cite{pollard1992theory, bardeen1995selective, sagar2008size, ref:rossi}. 

To retrieve the significant parameters of the response, we perform a global fit (GF) analysis of the $\Delta$A signal. Thirteen temporal traces in the 3.60 - 4.15 eV range of the $\Delta$A map are fitted simultaneously by imposing the same time constants. A satisfactory fit up to 1 ns can only be obtained using four exponential functions and two damped sinusoidal terms convoluted with a Gaussian response accounting for the pump pulse shape.
%\begin{equation}
%f(t) = I(t) * \sum_{i=1}^4 A_i e^{-t/\tau_i} + \sum_{j=1}^2 C_j \sin(2\pi\nu_j t + \phi_j) e^{-t/\tau_{D,j}}.
%\label{Fit_TiO2}
%\end{equation}
The results of the GF are shown as solid lines in Fig. 3(a), overlapped to the experimental data (displayed as dots). We remark that two damped sinusoidal terms are necessary to correctly reproduce the dynamics instead of one. While the first damped sinusoidal function is related to the large-amplitude oscillation that emerges in the slow recovery of the negative $\Delta$A signal, the second one accounts for an additional modulation (with smaller amplitude) that becomes apparent in the residuals of the fit when this term is not included. The frequencies of the two oscillations are imposed to be the same for all photon energies, since phenomenologically they do not display significant variations over the whole spectral range (Fig. 3(a)), and are found to be $\nu_1$ = 290 GHz and $\nu_2$ = 620 GHz. The GF analysis allows us to disentangle the spectral dependence of the two coherent excitations. In Fig. 3(b), we observe that both collective modes display the largest amplitude around 3.95 eV. Hence, the effect of these modes is to produce a stronger amplitude renormalization of the optical spectrum around 3.95 eV (i.e. at slightly higher energies than the exciton peak). The oscillation damping time (2.28 ps) is found to be constant over the probed spectral range, indicating that the lifetime is not governed by the optical properties (as the skin depth is much larger than the NP radius) but more by the intrinsic phonon damping. More importantly, within our accuracy, the phases of oscillations do not vary with photon energy (Fig. 3(c)). This aspect is of pivotal importance, as it sheds light on the detection mechanism of the oscillations in the vicinity of the exciton peak, as further described later.

The frequencies of the oscillatory components are very low compared to those of the optical phonons in anatase TiO$_2$ \cite{gonzalez1997infrared}. Here we demonstrate that they are indeed signatures of coherent acoustic phonons. In bulk single crystals, the frequency of coherent acoustic phonons varies with the probe photon energy \cite{ref:brillouin, ref:thomsen_prb}. In contrast, in nanosized objects (such as thin films or nanostructures), the confinement of the coherent wavepackets inside the structure leads the mode frequency to be governed by the boundary conditions. In our experiments, the frequency of the coherent oscillations does not vary with the probe photon energy, which suggests a confined nature for the acoustic modes inside the NPs, consistent with the usual assignement \cite{juve2010probing, ayouch2012elasticity}. To verify this scenario, we calculate the eigenfrequencies of the modes supported by a spherical anatase TiO$_2$ NP through Lamb's theory~\cite{ref:lamb}. This theory relies on the treatment of the NP as a homogeneous elastic sphere embedded in an infinite elastic medium, neglecting the anisotropy of the elastic constants of the crystal and the presence of the surrounding environment. This approach is justified for the relatively large NPs investigated here. This allows the assignment of the two coherent acoustic phonons as the fundamental spheroidal radial mode and its first overtone (i.e. the two first harmonics), with excellent agreement between the computed and experimental values (see Supporting Information). The fundamental spheroidal radial mode is typically known as ``radial breathing mode", since it involves a change of volume of the NP, leading to its contraction and expansion.

\begin{figure*}[tb]
	\begin{center}
		\centering
		\includegraphics[width=1\columnwidth]{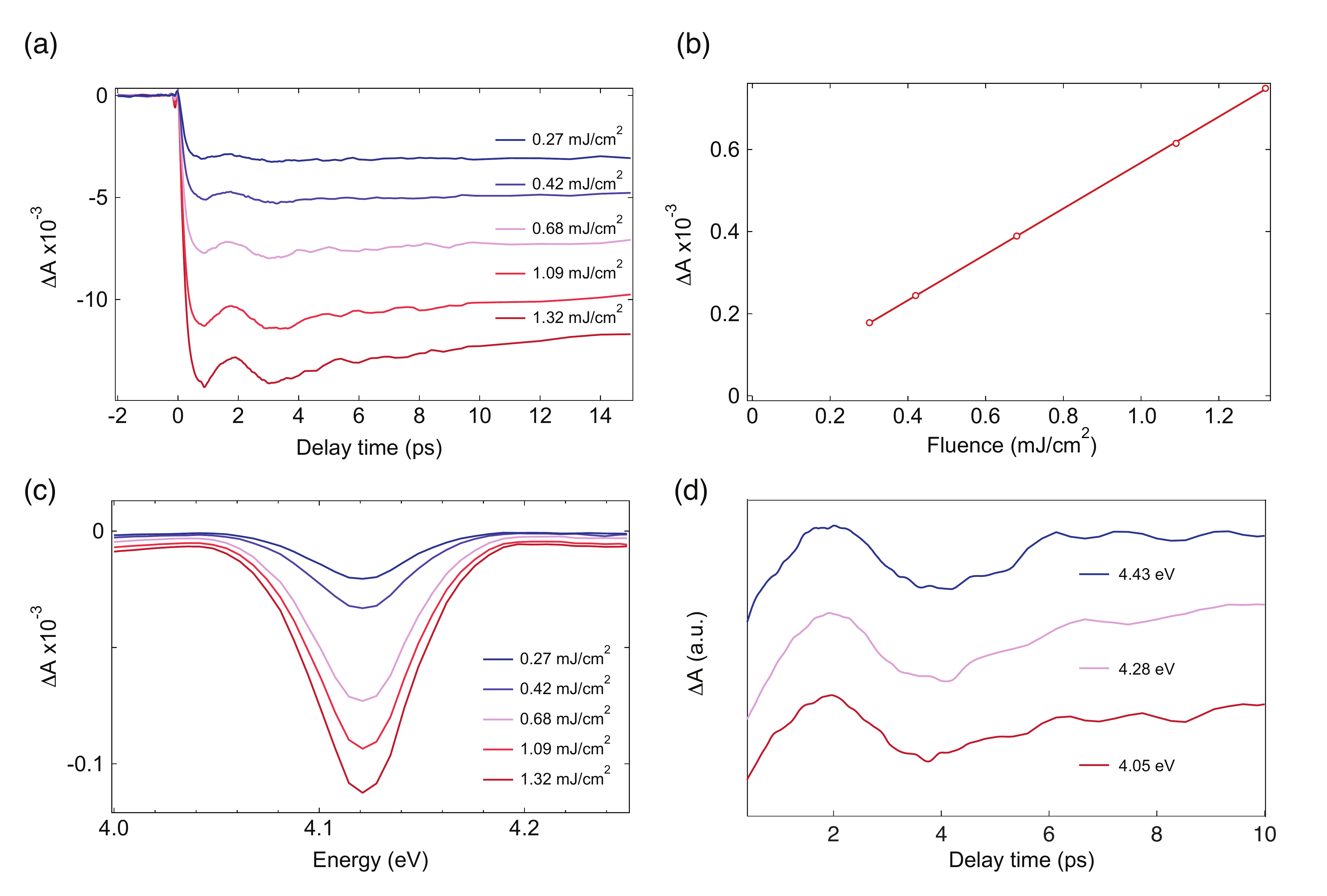}
		\caption{(a) Experimental temporal traces of $\Delta$A at 3.88 eV for different values of absorbed pump fluence. (b) Fluence dependence of the coherent acoustic phonon amplitude.  (c) $\Delta$A spectra at the maximum coherent acoustic phonon response as a function of absorbed pump fluence. (d) Pump photon energy dependence of the coherent acoustic phonon signal. The three curves have been shifted vertically for clarity.}
		\label{fig:AcPh_DAS}
	\end{center}
\end{figure*}

In order to gain further insight into the mechanism of coherent acoustic phonon generation and detection, we also explored the effect of the excitation density at 4.05 eV (Fig. 4(a)). We observe that the amplitude of the modulation at 3.88 eV scales linearly with increasing fluence (Fig. 4(b)). Depending on the probe photon energy, the modulation depth induced by the coherent acoustic phonons is 4-30\% of the total $\Delta$A at all fluences, which is larger than the changes observed in metallic NPs when the detection is optimized close to the plasmon resonance \cite{crut2015acoustic}. The oscillation damping is independent of the excitation density, indicating that the process behind the dissipation of the acoustic energy solely depends on the particle size distribution and the surrounding environment. We also track the time evolution of the exciton peak energy at the different fluences (Fig. 4(c)), finding no sizeable shift of the resonance within our experimental accuracy. Also the frequency and the phase of the breathing mode do not vary with fluence, confirming the absence of strong nonlinear effects or the influence of the carrier cooling dynamics on the coherent phonon signal \cite{bothschafter2013ultrafast}. Hence, these results demonstrate a surprising robustness of the exciton to the coherent strain pulse within the explored fluence range, confirming the validity of the same detection mechanism for all fluences and highlighting its linear response.

We finally address the microscopic origin of the generation and detection mechanisms for the observed coherent acoustic phonons. While the generation process reveals the coupling between the photoexcited uncorrelated carriers and the acoustic lattice modes, the detection mechanism offers insightful information on the impact of the coherent strain field on the excitonic resonances. Importantly, these two processes involve distinct bands and electron-hole distributions and, as such, may be governed by very different microscopic mechanisms.

The generation mechanism of coherent acoustic phonons in non-magnetic and non-piezoelectric materials is based on the competition between the non-thermal deformation potential (DP) coupling (electronic pressure) and the thermoelastic (TE) coupling (phononic pressure) \cite{ref:ruello, ref:thomsen_prb}. Whether the generation mechanism is governed by one or the other depends on a critical parameter: the ratio between the DP- and TE coupling contributions to the photoinduced stress ($\sigma_{DP}/\sigma_{TE}$). If this parameter is $\ll$ 1, the lattice contribution takes control of the coherent motion. In simple band insulators and semiconductors, the DP mechanism provides the main contribution. In the case of anatase TiO$_2$, due to the strong electron-phonon coupling, the photocarriers undergo an extremely fast intraband relaxation towards the edges of the respective bands ($\ll$ 50 fs) \cite{ref:baldini_excitonbleach}, suggesting that the phononic pressure can in reality compete with the electronic pressure for the coherent acoustic phonon excitation process. To quantify the electronic contribution to the photoinduced stress ($\sigma_{DP}$), we rely on the following expression \cite{ref:ruello}
\begin{equation}
\sigma_{DP} = \sum_k \delta N(k) \frac{dE_k}{d\eta},
\label{sigma_DP}
\end{equation}
\noindent where $\delta N$($k$) is the change of the electronic density at level \textit{k} and d$E_{k}$/d$\eta$ is the deformation potential parameter. Given that after 50 fs (\textit{i.e.} a much faster timescale than the detected coherent acoustic phonon period) the carriers have thermalized to the bottom of the respective bands at $\Gamma$ and $\sim$ X \cite{ref:baldini_excitonbleach}, the expression can be simplified as
\begin{equation}
\sigma_{DP} = -N B \Bigg[\frac{dE_e}{dP}\bigg|_{\Gamma} + \frac{dE_h}{dP}\bigg|_{X}\Bigg] = - N B (d_e + d_h),
\label{sigma_DP_2}
\end{equation}
\noindent where $N$ is the photoinduced carrier concentration and $B$ is the bulk modulus. 

The phononic contribution to the photoinduced stress ($\sigma_{TE}$) can be written as
\begin{equation}
\sigma_{TE} = -\alpha_V B \Delta T_L = -\alpha_V B N E_{exc}/C_L,
\label{sigma_TE}
\end{equation}
\noindent where $\alpha_V$ is the volumetric thermal expansion coefficient. For a tetragonal crystal, $\alpha_V = 2\alpha_{\perp} + \alpha_{\parallel}$, where $\alpha_{\perp}$ and $\alpha_{\parallel}$ are the in-plane and  the out-of-plane thermal expansion coefficients, respectively. $\Delta T_L$ is the lattice temperature, $C_L$ is the lattice heat capacity per unit volume and $E_{exc}$ is the excess energy with respect to the optical bandgap energy. The ratio $\sigma_{DP}$/$\sigma_{TE}$ reads
\begin{equation}
\frac{\sigma_{DP}}{\sigma_{TE}} = \frac{C_L (d_e + d_h)}{\alpha_V  E_{exc}}.
\label{sigma_DP_TE}
\end{equation}

To evaluate this expression, we calculate the electron and hole DPs from \textit{ab initio} calculations. Specifically, we compute the single-particle excitation spectrum of anatase TiO$_2$ at the GW level of theory for different strengths of the unit cell isotropic deformation along the three crystal axes. This hydrostatic pressure is introduced to mimic the displacement that the unit cell undergoes during the occurrence of the fundamental breathing mode. The GW treatment yields the DPs $d_e$ = -0.089 eV/GPa and $d_h$ = -0.065 eV/GPa. This leads to $\sigma_{DP}$/$\sigma_{TE}$ = -27.47, which indicates that the DP mechanism provides the dominant contribution to the generation of the observed confined coherent acoustic phonons. This conclusion is also confirmed experimentally by repeating our measurements at different pump photon energies while keeping the absorbed excitation density fixed (Fig. 4(d)). Indeed, increasing the excess energy of the photoexcited charge carriers leads to an increase in the energy stored in the lattice once the carriers have relaxed to the band edges (intraband relaxation process). Different lattice temperatures would give rise to distinct coherent acoustic phonon amplitudes in the pump-probe traces. Our experimental data rule this possibility out, as the coherent acoustic phonon amplitude is independent of the carrier excess energy. As such, our combined theoretical-experimental approach sheds light on the generation process of the confined acoustic modes. We underline that these results can be obtained only with a reliable estimate of the DPs from many-body perturbation theory. In contrast, by relying on simple density-functional theory \cite{ref:yin_effective}, the value $\sigma_{DP}$/$\sigma_{TE}$ = -0.84 is obtained, implying a larger contribution of the TE coupling to the photoinduced stress. In conclusion, our analysis demonstrates the crucial role of many-body perturbation theory in providing the correct interpretation of the coherent acoustic phonon generation mechanism in this wide-bandgap insulator.

The detection process of the coherent acoustic phonons represents the most intriguing aspect of our study. In general, the use of a broadband probe in ultrafast optical spectroscopy reveals information on the response of interband transitions and excitonic resonances during the mechanical motion induced by coherent phonons. Indeed, the phase shifts in the oscillations based on the probe photon energy can yield assignment of \textit{energy} versus \textit{amplitude} modulation of a resonance \cite{sagar2008size}. The \textit{energy modulation} scenario results in a phase inversion of $\pi$ below and above the resonance, and the intensity of the coherent oscillation detected in the pump-probe signal is roughly proportional to the square of the derivative of the absorption spectrum~\cite{sagar2008size, sanders2013theory}. This is the typical condition observed for most phonon-exciton coupling phenomena in the time-domain \cite{kirschner2017phonon}. In contrast, in the less conventional \textit{amplitude modulation} scenario, it is the intensity of the resonance in the absorption spectrum that varies with time while the peak energy is fixed. As a result, the coherent phonon phase remains constant over the resonance. The microscopic origin behind these effects is related to the particular parameter of the electronic Hamiltonian that is modulated by the oscillating ionic motion. In the case of an \textit{energy modulation}, the coherent lattice motion modulates the band structure via the DP mechanism, which results in a periodic energy renormalization of the excitonic and the interband charge excitations. In the much less frequent \textit{amplitude modulation}, the ionic motion acts directly on other parameters of the electronic Hamiltonian, such as: i) The hopping integral of an electron from site to site, modifying the electron bandwidth and therefore the density of states, and/or; ii) The transition dipole moment of an interband transition, thus creating a Herzberg-Teller wavepacket~\cite{ref:chen_Fe2O3, ref:ishii}. The modification of the electronic density of states appears as a more likely explanation in the case of anatase TiO$_2$. Indeed, excitons are of the charge-transfer type in this material, being formed by holes in oxygen 2$p$ states and electrons in titanium 3$d$ states. During the expansion and contraction of the NPs induced by the acoustic breathing mode, the Ti-O distances are periodically modified, leading to a modulation of the overlap between the relevant orbitals contributing to the many-body exciton state. As a result, the hopping integral between the two lattice sites and the electronic density of states are readily affected. To our knowledge, this effect has never been directly observed and selectively controlled.

To conclude, our experimental data clearly show an amplitude modulation of the resonance during the coherent ionic motion. The modulation is caused by the breathing mode of the NP and its first overtone, which are shown by many-body perturbation theory to be generated from the deformation potential coupling. At this stage we cannot distinguish whether the pronounced amplitude modulation is produced by a variation of density of states, a change in the transition dipole moment of the excitonic transition or both, but the overall result is remarkable and unique. It also offers the possibility of applications in the field of sensors, especially considering that anatase TiO$_2$ is a cheap material and is already embedded in a large range of devices. In this respect, future efforts to increase the coherence time of the confined wavepackets, e.g. by using monodisperse ensembles of NPs and embedding them in an acoustic impedance-mismatched matrix, will allow to achieve long-lasting modulations of the exciton oscillator strength.

\section{Methods}
The TiO$_2$ NPs were prepared using the sol-gel method~\cite{ref:mahshid}. The synthesis was carried out in a glove box under argon atmosphere. Titanium isopropoxide (Sigma Aldrich, 99.999$\%$ purity) was used as precursor and was mixed with 10 ml of 2-propanol. This mixture was added dropwise under vigorous stirring to cold acidic water ($2\,^{\circ}\mathrm{C}$, 250 ml H$_2$O, 18 M$\Omega$, mixed with 80 ml glacial acetic acid, final pH 2). At the beginning the mixture looked turbid, but after stirring it in an ice bath for 12 hours, it became transparent as the amorphous NPs were formed. Half of the mixture was left stirring for days to stabilize the NPs. The other half was peptized at $80\,^{\circ}\mathrm{C}$ for about 2 hours until the liquid turned into a transparent gel. The gel was autoclaved at $230\,^{\circ}\mathrm{C}$ for 12 hours. During this process the previous amorphous TiO$_2$ sample became denser and underwent a phase transition, resulting in anatase TiO$_2$ NPs. After the autoclave, the NPs have precipitated to the bottom of the container. They were separated from the supernatant and added to 100 ml acidic water (pH 2) to obtain a white colloidal solution with a final concentration of ca. 5 g/L. In Ref. \cite{ref:rittmann-frank, santomauro2015femtosecond}, we reported the details of the sample characterization by means of X-ray diffraction and transmission electron microscopy. Using these techniques, the good quality of the anatase phase and the spherical shape (with an average diameter of approximately 25 nm) of the NPs were demonstrated. Furthermore, the room temperature steady-state absorption spectum of the anatase TiO$_2$ NPs was measured and shown in Ref. \cite{ref:baldini_TiO2}.

The ultrafast broadband deep-ultraviolet spectroscopy measurements have been performed using a set-up described in detail in Ref. \cite{ref:aubock}. Briefly, a 20 kHz Ti:Sapphire regenerative amplifier (KMLabs, Halcyon + Wyvern500), providing pulses at 800 nm (1.55 eV), with typically 0.6 mJ energy and around 50 fs duration, pumps a non-collinear optical parametric amplifier (NOPA) (TOPAS white - Light Conversion) to generate sub-90 fs visible pulses (1.77 - 2.30 eV range). The typical output energy per pulse is 13 $\mu$J. Around 60$\%$ of the output of the NOPA is used to generate the narrowband pump pulses. The visible beam, after passing through a chopper, operating at 10 kHz and phase-locked to the laser system, can be focused onto a 2 mm thick BBO crystal for obtaining the UV pump pulse. In this case, the pump photon energy is controlled by the rotation of the crystal around the ordinary axis and can be tuned in a spectral range up to $\geq$ 0.9 eV ($\geq$ 60 nm) wide. The typical pump bandwidth is 0.02 eV (1.5 nm) and the maximum excitation energy is about 120 nJ. The pump power is recorded on a shot-to-shot basis by a calibrated photodiode for each pump photon energy, allowing for the normalization of the data for the pump power. The remaining NOPA output is used to generate the broadband UV probe pulses with $\geq$ 100 nm bandwidth through an achromatic doubling scheme. Pump and probe pulses, which have the same polarization, are focused onto the sample, where they are spatially and temporally overlapped. The typical spot size of the pump and the probe are 100 $\times$ 150 $\mu$m$^2$ and 40 $\times$ 44 $\mu$m$^2$ full-width half-maximum respectively, resulting in a homogeneous illumination of the probed region.

The colloidal solution circulated into a 0.2 mm thick quartz flow-cell to prevent photo-damage and its concentration was adjusted to provide an optical density of approximately 0.4 at the pump photon energy of 4.05 eV. The probe was measured after its transmission through the sample and its detection synchronized with the laser repetition rate. The difference of the probe absorption with and without the pump pulse has been measured at varying time delays between the pump and the probe, thanks to a motorized delay line in the probe path. After the sample, the transmitted broadband probe beam was focused in a multi-mode optical fiber (100 $\mu$m), coupled to the entrance slit of a 0.25 m imaging spectrograph (Chromex 250is). The beam was dispersed by a 150 gr/mm holographic grating and imaged onto a multichannel detector consisting of a 512 pixel CMOS linear sensor (Hamamatsu S11105, 12.5 $\times$ 250 $\mu$m pixel size) with up to 50 MHz pixel readout, so the maximum read-out rate per spectrum (almost 100 kHz) allowed us to perform shot to-shot detection easily. The described experimental setup offered a time resolution of 150 fs, but this could be improved to 80 fs with the adoption of a prism compressor on the pump pulses, at the expenses of a reduction of the probe bandwidth in the achromatic frequency doubling scheme.

\section{Acknowledgement}
	
We acknowledge support by the Swiss NSF via the NCCR:MUST and PNR70: 407040\_154056, and by the European Research Council Advanced Grants DYNAMOX 695197 and QSped-NewMat. E.B. acknowledges support by the Swiss NSF under fellowship P2ELP2\_172290. 
	
\clearpage
\newpage


\section{S1. Estimation of the Carrier Density}

We calculate number of excited electron-hole pairs per unit cell of anatase TiO$_2$ by taking into account a sample of NPs, flowing in a quartz flow cell of thickness $l$ = 0.2 mm with a concentration of $c$ = 0.106 g/L and an optical density of $OD$ = 0.4. The sample is pumped with a repetition rate of $f$ = 20 kHz, average power $P$ at a photon energy of $E_{pump}$ = 4.05 eV). The energy per pulse is therefore $E_{pulse} = P/f$, corresponding to a number of photons per pulse of $N_{pulse} = E_{pulse}/E_{ph}$. The pump is focused on an area of $A_{foc} = \pi \times 50 \mu m \times 75 \mu m = 1.18 \times 10^{-8} m^2$, corresponding to a focusing volume of $V_{pump} = A_{foc} l = 2.36 \times 10^{-12}m^3$. Since the radius of one NP is $R$ $\sim$ 14 nm, the volume can be estimated by approximated the NP to a sphere. This yields $V_{np} = \frac{4}{3}\pi R^3 = 1.15 \times 10^{-23}m^3$. The anatase TiO$_2$ density is $\rho$ = 3.9 $10^6$g/$m^3$ and the mass of one NP is $M_{np} = V_{np} \rho = 4.5 \times  10^{-17}$g. Given the concentration $c$, the number of NPs in the volume where the pump is focused is $N_{foc} = c V_{pump} 10^3/ M_{np} = 5.57 \times 10^6$. The number of photons absorbed in a pulse is $N_{abs}=\alpha N_{pulse}$, where $\alpha = 10^{-OD} = 0.398$. In conclusion, the photoexcited carrier density $n$ is calculated as the ratio between total number of absorbed photons and the total illuminated volume $N_{foc} \times V_{pump}$.

\section{S2. Assignment of the Coherent Oscillations}

The confinement condition enables to identify the CAPs by calculating the eigenfrequencies of the modes supported by a spherical anatase TiO$_2$ NP through Lamb's theory of vibrations~\cite{ref:lamb}. This theory relies on the treatment of the NP as a homogeneous elastic sphere embedded in an infinite elastic medium, neglecting the anisotropy of the elastic constants of the crystal and the presence of the surrounding environment. This approach is justified for the relatively large NPs investigated here. Lamb's classical description can be exploited just to compute the approximate eigenfrequencies of the modes and it is not effective in the estimation of the damping associated to the oscillations. The solution of the Navier's equation for the displacement field with proper boundary conditions at the surface of the particle gives rise to two classes of eigenmodes, the spheroidal and the torsional ones. A general spheroidal oscillation involves both longitudinal and transverse character, corresponding to a movement accompanied by a volume change of the particle. On the contrary, torsional modes are completely transverse and do not influence the volume of the sphere. In low-frequency spontaneous Raman scattering both types of modes can be monitored, while in femtosecond pump-probe experiments only the spheroidal family has been observed. The symmetry group of the sphere allows the characterization of each mode in terms of an angular momentum number l, which gives an estimation of the number of wavelengths along a circle on the surface of the particle. There is a general agreement in considering modes with even-valued $l$ ($l$ = 0, 2, ...) as Raman-active, while those with odd-valued l ($l$ = 1, 3, ...) as infrared-active. The oscillation with $l$ = 0 can be regarded as a special type of spheroidal mode, since it is associated only with a radial displacement. This fundamental radial mode produces a variation in the size of the NPs while maintaining the shape unchanged, and yields the dominant response in time-resolved optical experiments. Another index, $n$, is used to determine the order of the mode, defining the fundamental oscillation ($n$ = 0) and its overtones ($n$ $>$ 1). 

According to Lamb's theory and its extension for large-enough NPs~\cite{ref:sakuma, ref:tamura}, the eigenfrequencies of the spheroidal modes can be derived as
\begin{equation}
\Omega_{l,n} = \xi_{l,n} \frac{v_{Lp}}{R},
\label{eigenfrequencies}
\end{equation}
where $v_{Lp}$ is the longitudinal sound velocity in the NP, $R$ is the radius of the sphere and $\xi_{l,n}$ are the normalized eigenfrequencies that can be found as a solution of the equation 
\begin{equation}
\tan(\xi_{0,n}) = \frac{\xi_{0,n}}{1- \frac{1}{4}\xi^{2}_{0,n} \left( \frac{v_{Lp}}{v_{Tp}}\right)^2 }.
\label{tan}
\end{equation}
To assign the mode associated with the oscillation observed in our data, we numerically solve Eq. \ref{tan}. We use the values of $v_{Lp}$ = 8880 m/s and $v_{Tp}$ = 3900 m/s for the sound velocities, which refer to the case of anatase TiO$_2$ NPs embedded in a glass matrix \cite{ref:ivanda}, and $R$ = 14 nm, which is evaluated by TEM measurements \cite{ref:rittmann-frank}. As a result, for n = 0 and n = 1, we find the eigenvalue $\xi_{0,0}$ = 2.85 and $\xi_{0,1}$ = 6.16 respectively. The substitution of this parameter into Eq. \ref{eigenfrequencies} leads to the calculation of the frequencies $\hbar\Omega_{0,0}$ = 1.20 meV and $\hbar\Omega_{0,1}$ = 2.56 meV, which are in excellent agreement with our experimental data. This allows the assignment of the two CAPs as the fundamental spheroidal radial mode and its first overtone. The fundamental spheroidal radial mode is typically known as ``breathing mode", since it involves a surface-localized change of volume of the NP, leading to its contraction and expansion. Further corrections should consider the size-distribution of the NPs around the central radius R, as well as the contribution given by their elliptical shape. The rapid dephasing of the oscillation can be further ascribed to the combination of the particle size distribution and the damping produced by the surrounding environment. 
\newline

\begin{table}[h!]
	\centering
	\normalsize
	\begin{tabular}{ccc}
		
		\textbf{Parameter} & \textbf{Value} \\
		Thermal expansion coefficients (K$^{-1}$) &\\
		$\alpha_a$ &		7.57 $\times$ 10$^{-6}$ ~~~\cite{ref:rao} \\
		$\alpha_c$ & 	3.66 $\times$ 10$^{-6}$ ~~~\cite{ref:rao} \\
		Deformation potentials (eV/GPa) &\\
		$d_e$ & 	-0.089\\
		$d_h$ & 	-0.065\\
		Excess energy $E_{exc}$ (eV) &  0.80 \\
		Heat capacity per mole $C_m$ (J mol$^{-1}$ K$^{-1}$) & 55.100 ~~~\cite{ref:smith_heat} \\
		Mass density $\rho$ (m$^{-3}$) & 3.89 $\times$ 10$^6$\\
		Molar mass $M$ (g mol$^{-1}$) & 79.9 \\
	\end{tabular}
	\caption{Parameters used for evaluating the DP and TE coupling contributions to the photoinduced stress in anatase TiO$_2$ NPs.}
	\label{tab_TiO2}	
\end{table}

\newpage


	\newpage

\providecommand{\noopsort}[1]{}\providecommand{\singleletter}[1]{#1}%
\providecommand{\latin}[1]{#1}
\providecommand*\mcitethebibliography{\thebibliography}
\csname @ifundefined\endcsname{endmcitethebibliography}
{\let\endmcitethebibliography\endthebibliography}{}

\end{document}